*Article*

# Multi-Feature Data Fusion-Based Load Forecasting of Electric Vehicle Charging Stations Using a Deep Learning Model

Prince Aduama [1,*], Zhibo Zhang [1] and Ameena S. Al-Sumaiti [2,*]

1. Department of Electrical Engineering and Computer Science, Khalifa University of Science and Technology, Shakhbout Bin Sultan St Zone 1, Abu Dhabi 127788, United Arab Emirates
2. Advanced Power and Energy Center, Department of Electrical Engineering and Computer Science, Khalifa University of Science and Technology, Abu Dhabi 127788, United Arab Emirates
* Correspondence: 100060989@ku.ac.ae (P.A.); ameena.alsumaiti@ku.ac.ae (A.S.A.-S.)

**Abstract:** We propose a forecasting technique based on multi-feature data fusion to enhance the accuracy of an electric vehicle (EV) charging station load forecasting deep-learning model. The proposed method uses multi-feature inputs based on observations of historical weather (wind speed, temperature, and humidity) data as multiple inputs to a Long Short-Term Memory (LSTM) model to achieve a robust prediction of charging loads. Weather conditions are significant influencers of the behavior of EV drivers and their driving patterns. These behavioral and driving patterns affect the charging patterns of the drivers. Rather than one prediction (step, model, or variables) made by conventional LSTM models, three charging load (energy demand) predictions of EVs were made depending on different multi-feature inputs. Data fusion was used to combine and optimize the different charging load prediction results. The performance of the final implemented model was evaluated by the mean absolute prediction error of the forecast. The implemented model had a prediction error of 3.29%. This prediction error was lower than initial prediction results by the LSTM model. The numerical results indicate an improvement in the performance of the EV load forecast, indicating that the proposed model could be used to optimize and improve EV load forecasts for electric vehicle charging stations to meet the energy requirements of EVs.

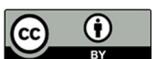

## 1. Introduction

Recent initiatives to cut greenhouse gas emissions to combat climate change have led to increasing renewable energy penetration into modern grids [1–3]. Many nations are planning to phase out automobiles powered by internal combustion engines (ICEs) [2,3]. This has sparked a push to boost electric vehicle (EV) penetration into the modern vehicular fleet [4–7]. There are three main reasons for the recent adoption of EVs: affordability, increasing oil prices, and development sustainability [8,9]. In relation to affordability, for example, battery costs fell drastically from almost $1000 per kWh in 2010 to less than $156 per kWh in 2019 [10]. An International Energy Agency (IEA) report estimated that the transportation sector could contribute to reduced $CO_2$ emissions by up to 21% by 2050 [11], and electric vehicles would be a key factor in achieving this. The Electric Power Research Institute [12] estimates that for low, moderate and high penetration scenarios, EVs would represent 20, 62, and 80%, respectively of all vehicles.

However, EV proliferation comes with a number of issues, including battery capacities and range anxiety related to EVs. EVs are means of energy storage [13], but range anxiety which is concerned with lack of the EV's energy storage capacity to traverse the distance required to reach the destination, is still a major problem. A survey conducted by Consumer Reports indicated that 49% of a total of 3323 respondents wanted at least a 300-mile charging range for their EVs, and a lot more (about 63%), expected a minimum of 250 miles of range [14]. Such an issue is considered one of the most significant psychological impediments to widespread public adoption of EVs. Because of this, the electrical demand for EVs is more uncertain than that for residential energy need, and is seen in the variety of EV travel patterns and range anxiety. In addition, increased penetration and uncertainty of EVs may have several effects on the power system when the charging stations are connected to the grid. Effects of EV charging/discharging on the energy grid have been investigated previously [15,16]. Charging and discharging of EVs could cause imbalances between power supply and demand, and could lead to power quality and stability issues [17,18].

From the viewpoint of the power system, EVs can influence power system flexibility [19]. PEV charging has a random pattern depending on drivers, and demand can coincide with peak loads. Since this is uncontrollable, it may result in an uncertain load profile. Therefore, aggregators have been introduced in the power sector [20,21]. In [2], information on the demand charge of PEVs is provided. In addition, strategies that assure profitability are considered, and rewording for PEV drivers is discussed. An important factor influencing the model in [2] is the accuracy of forecasting the PEV's demand, as discussed in [3]. It is important to estimate the aggregated EV charging station demand on a power system to aid in power generation planning to keep up with load increases as EVs acquire more market dominance.

Regarding future smart grids, establishing efficient charging infrastructure control and scheduling schemes is essential to accommodate more clean energy, reduce carbon emissions, and alleviate peak charging loads. The type of data used in the prediction process could also affect the accuracy of the results generated. Majidpour et al. in [22] forecasted EV charging loads considering various datasets, such as charging records (customer profile) and charging station (outlet measurement) records. The authors concluded that the charging record is more accurate than the outlet measurement record after four different prediction algorithms were used to test the model. The research in [23] presents a short-term forecasting model for PHEV aggregate loads. A mathematical model was used to observe the features of real-world data, which was then used for forecasting. However, a theoretically infinite number of vehicles were considered, and therefore the model is not suitable for a smaller EV vehicular fleet. To predict the sales of plug-in EVs, Duan et al. [24] used customer preferences determined by historical data between hybrid EVs and internal combustion engine (ICE) automobiles. The forecast model was based on the assumption that factors related to vehicular choice between EVs and ICEs would shift from convenience to cost. In [25], statistical data of travel patterns of electric buses, together with a backdrop propagation neural network, was involved in the prediction of the number of electric buses needed for battery swapping at hourly time intervals. The authors also considered forecasting uncertainties using a modification of the kernel density estimator.

Artificial intelligence (AI) techniques have been employed in several aspects of EV applications. These include load demand and battery capacity prediction [26], charging and discharging [27,28], and energy management [29,30]. In [4], an overview of the growth of AI technologies and how they have been applied to EVs is presented. The authors provide a good perspective on the most popular AI techniques and how new techniques are gaining more prominence in their application to EVs. Much research work involving AI related to EVs has been geared towards optimizing user experience and enhancing the resilience of the power grid to the load patterns of EVs. The authors in [6] employed a transformer-based deep learning model to classify various topics among EV users, and reported a high accuracy greater than 91%. The only downside of this model was its training and testing time of 1 to 4 hours, compared to 1 to 90 min for CNN and LSTM models. The study in [7] examined how adaptable and capable of regulation EV aggregators are in participating in electric power markets. Demand response and spot market dynamic pricing strategy were solved using the DDPG reinforcement learning algorithm to maximize transaction revenue. In [31], a new hybrid classification approach considering RNN and LSTM networks is proposed. An unsupervised classifier was first utilized to find the hidden travel patterns in the historical PEVs data. The PEV data were then classified into a cluster-specific forecasting network by a supervised classifier. Deep learning forecasting networks were implemented according to LSTM networks to investigate the LSTM features of PEV behaviors. Although, the model brought about improvements in generating realistic travel patterns, the model used only the departure time for its classification, which could affect its efficiency.

Other prediction models using AI techniques have been implemented within several areas of electric vehicle research. In [32], a graph convolution network model was implemented to determine the level of use of charging stations within a particular period. In [33], historical charging data were used to forecast future charging load. This forecasting was done using a variant of LSTM known as the gated recurrent unit (GRU) model. Other prediction models have focused on the batteries of EVs. Forecasts for the end of battery life are discussed in [34]. A hybrid approach was utilized for this prediction. First, the authors employed empirical mode decomposition (EMD) to break down the various time indices of the forecast. Then, a sequential importance sampling (SIC) was used in implementing a particle filter for the estimation process. Some research has focused on the range prediction of EVs. The authors in [35] present a review of electric vehicle range prediction. Factors that affect the range of the EV, such as the vehicle design, environment, and human factors are thoroughly discussed. Furthermore, the authors present methods and approaches that are implemented for range prediction of EVs. In [36] data on EV trips were utilized in models for the prediction of EV energy consumption. The authors combined three different base models, K-Nearest Neighbor, Random Forest, and Decision Tree, to form an Ensemble Stack Generalization approach for prediction to improve the results offered by each base model. A comparative analysis between several data-driven machine learning and statistical techniques for battery state of charge estimation is presented in [37].

One of the ways of balancing the power system to reduce power quality and stability issues is to estimate the energy charging demand of electric vehicles within a certain period (e.g., for one day), and to balance that load. Of course, the exact load would be almost impossible to ascertain. The load forecast for charging stations should be as close as possible to the actual demand. Much research has been done in this area to address this problem. A mixed integer non-linear programming (MINLP) approach was used in [38] to find the best location and number of fast charging stations. A genetic algorithm technique was applied to reduce the development costs of the station, the costs associated with EV users, and the costs associated with grid operators. However, as noted in [9], this approach does not consider terrain features, available space and weather parameters, all of which could affect the siting of stations as well as their energy consumption. Furthermore, the model had a very high computational time of 420 min.

A deep reinforcement learning-based model was developed in [2] to cut down on origin-to-destination distance and curtail the total time of charging of EVs. The model included a flexible reward function to balance the charging time and the origin-destination distance reduction of EVs. However, even though the model improved upon the traditional forecasting model with a shorter average charging time, the actual traffic and charging station status data were not taken into consideration. In [39], a reinforcement learning strategy was used to improve charge scheduling and pricing issues. The proposed algorithm's decisions regarding charging and pricing each time were based solely on the observation of prior events. However, in a reinforcement learning problem, the continuous state and action vary with time.

A novel forecasting technique based on a data fusion and deep learning technique is presented in this study to provide a possible solution for the deficiencies in LSTM model prediction and to improve on the accuracy of EV charging load (energy demand) prediction. This study uses real-world data about local weather conditions, including temperature, humidity, and wind speed as training data to develop a forecasting model of EV energy demand, which results in the charging stations' energy demand. The general overview of the approach is shown in Figure 1. To the best of the authors' knowledge, the data fusion approach has not been implemented to date for EV load forecasting.

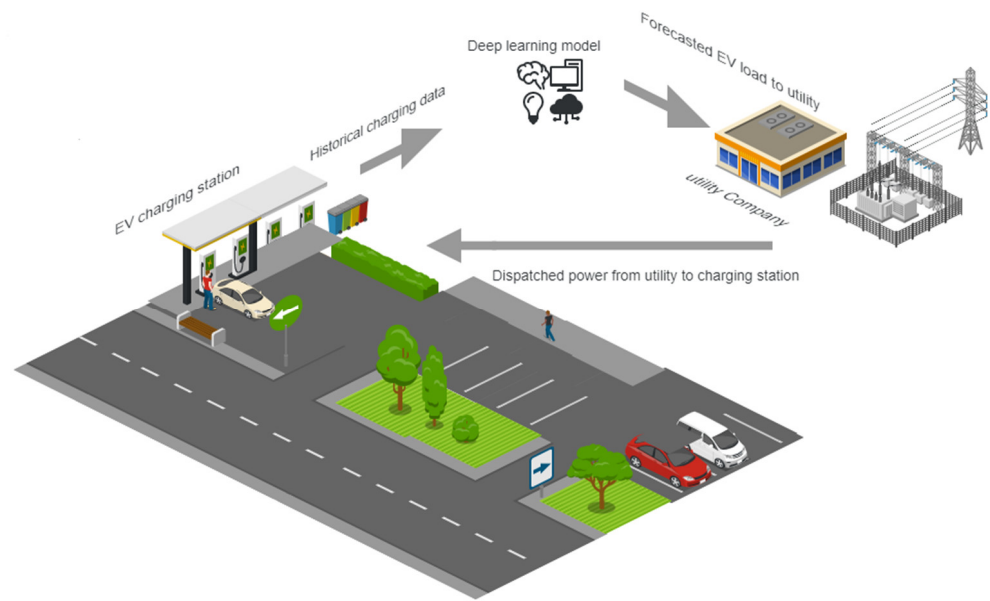

**Figure 1.** General overview of the forecast model.

This paper is organized as follows: the proposed data fusion model is introduced in Section 2; the forecasting approach being implemented is presented in Section 3; the results of the study are discussed in Section 4, and, the conclusion is presented in Section 5.

**2. System Modelling**

The system model consists of two key components; (1) the deep leaning (LSTM) model, which implements the initial prediction of the EV loads, and (2) the data fusion model, which optimizes the initial prediction by the LSTM model. This section has two parts. First, the various equations relative to the deep learning model are presented. Furthermore, the relation of these equations to the prediction model is explained and discussed thoroughly. The second part of this section discusses details of the data fusion model that are used to optimize the forecast of the deep learning model. The algorithm for the data fusion model is presented.

*2.1. LSTM Model*

Deep learning is an effective technique for studying large-dimensional issues with intricate relationships, such as time series forecasting, pattern recognition in video and picture data, and audio processing [17,40,41]. Deep learning concepts have excellent capacity to derive the key features of a large phenomenon based on past data. This has a great advantage over other data driven techniques [42].

Recurrent Neural Network (RNN) models are improved versions of neural networks that can exploit the features of previous information. RNN models are able to exploit sequential structure in data and are commonly used in time series predictions [43]. However, RNN networks have deficiencies in learning and storing information for long periods [17] and the RNN variant; the long-short term memory (LSTM) model has proven to be better in correcting these deficiencies. As a result, the LSTM model, which is a derivative of the RNN, was selected for this forecasting. This was due to its benefit in overcoming the gradient norm's sharp decrease for the long-term component within the model [31]. These models are effective in capturing the short-term variations in energy demand needed for forecasting [44]. In addition, the long-term trends in the charging patterns can be captured using the LSTM model.

The collected data are sorted and classified. The features to be captured are the energy demand, temperature, humidity and wind speed. The LSTM model is then trained using the data to extract the important model characteristics. The LSTM block is shown in Figure 2. The block has three operation gates, the input, output and forget gates. The networks are constructed from stacks of several LSTM blocks. With this structure, the output of a particular network at state $s-1$ and the preceding network output at state $s$ serve as the input data for every LSTM block of state $s$ [31].

In the model, the hidden features to be extracted, such as the temperature and wind speed, are propagated through different LSTM blocks during training. This increases the precision of the learning process. Equations (1)–(5) provide the general equations for the LSTM block in Figure 2. The supervised classification approach is shown in Algorithm 1.

$$i_s^l = \sigma(Wi^l S_S^{l-1}) + Whi^l S_{S-1} + bi^l \tag{1}$$

$$f_s^l = \sigma(Wi\varphi^l S_S^{l-1}) + Whi^l S_{S-1} + bf^l \tag{2}$$

$$c_s^l = f_s^l c_{s-1}^l + i_s^l \tanh(Wi\gamma^l S_s^{l-1} + Wh\gamma^l S_{l-1}^l + bc^l) \tag{3}$$

$$S_S^l = o_s^l \tanh(c_s^l) \tag{4}$$

$$o_s^l = \sigma(Wio^l S_S^{l-1}) + Who^l S_{S-1} + bo^l \tag{5}$$

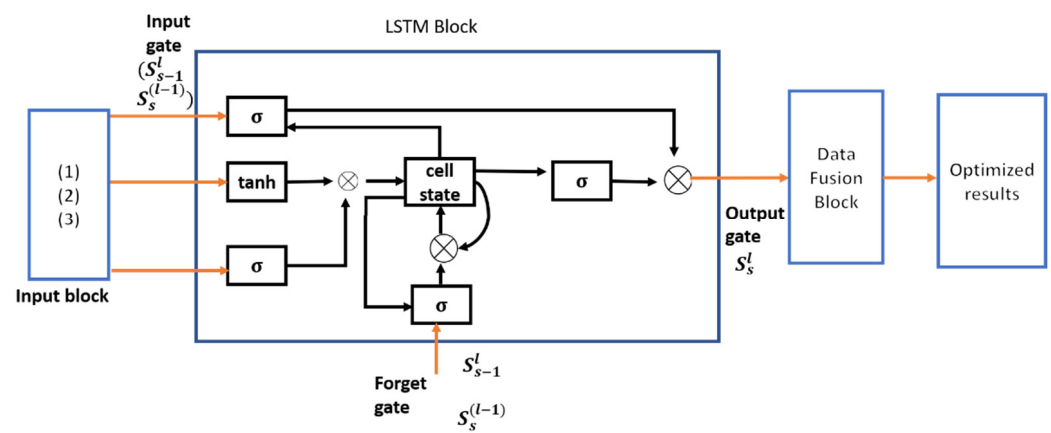

**Figure 2.** Block Diagram of system input, LSTM model and data fusion model.

---

**Algorithm 1** LSTM deep learning classification.

1. **Start Operation**

Sort each trip according to data on temperature, humidity, and wind speed, then assign a certain forecasting network to each cluster.
**Input**: Temperature, humidity, wind speed.
**Target**: Centroid of Clusters

2. Match each input sample of temperature, humidity, and wind speed to its cluster using a deep LSTM network with the aim being the centroid of each cluster.
3. **End operation**

---

*2.2. Data Fusion Method*

Data fusion is the process of combining various data sources to create information that is more reliable, accurate, and practical than information that can be obtained from a single data source alone. The capacity of humans and animals to combine information from numerous senses to improve their survival gave rise to the concept of data fusion [45]. A fusion of sight, touch, smell, and taste, for example, might indicate whether or not a thing is edible. Three sorts of data fusion approaches exist: data association, state estimation, and decision fusion [45]. Our research mainly implements the Dempster-Shafer evidence theory (D-S theory) [46], a decision fusion approach of data fusion techniques to optimize the EV charging demand forecasting using the LSTM technique.

D-S theory is a theory for dealing with situations involving uncertainty. The use of "interval estimate" rather than "point estimation" for the representation of uncertain information is the most important feature of D-S theory, which has, remarkable flexibility in defining unknowns and uncertainty and in properly capturing evidence. D-S theory can emphasize not only the objectivity of things but also the subjectivity of human estimation. D-S theory requires the classification of four fundamental ideas: frames of discernment, basic probability assignment (BPA), belief function, and plausibility function [47]. The D-S theory presupposes that the sets of items that make up the frames of discernment, represented by U, are exhaustible and mutually exclusive, and is given by the equation:

$$U = \{u_1, u_2, \ldots, u_n\} \tag{6}$$

Note that a set of size **n** including itself has exactly $2^n$ subsets that defines the power set, denoted as $2^U$ in D-S theory. Furthermore, there is a one-to-one correspondence between $2^U$ and the correct answers to all possible questions corresponding to the environment [47,48]. For example, for:

$$U = \{A, B, C\} \tag{7}$$

The corresponding $2^U$ is:

$$2^U = \{\emptyset, \{A\}, \{B\}, \{C\}, \{A,B\}, \{A,C\}, \{B,C\}, \{A,B,C\}\} \tag{8}$$

In our research, the possible prediction results for the EV load forecast can be regarded as possible inputs A, B, and C in the D-S theory, and the corresponding power subset in our load prediction instances denotes the average of inputs, which makes the prediction system more robust. Furthermore, there is no subset component $\emptyset$ in this model, and therefore, for the three different electrical vehicle load predictions in this model there would be only seven subset components. For instance, the component subset {A, B} would be an average of the original prediction A and B.

The second concept in D-S theory is BPA. In D-S theory, it is customary to consider trust of evidence similar to the quality of a physical object, i.e., the quality (mass) of the evidence supports a trust. Each mass can be formalized as a function that maps each element of the power set to a real number within the interval [0, 1], [49]. The function is described formally as:

$$m : 2^U \rightarrow [0, 1] \tag{9}$$

The term "mass" is derived from comparisons between the actual charging load data and the predicted loads. The quantity of mass in our studies increases as the similarity between the realistic loading data and the prediction data increases. Ideally, if the predicted load data and the real load data are exactly equivalent to each other, the mass for this prediction would be 1. The mass of an empty set is usually defined as 0, and the mass sum of all subsets of the power set $2^U$ of U is 1.

$$\sum_{X \in 2^U} m(X) = 1 \tag{10}$$

Table 1 illustrates that mass has a much larger degree of freedom than probability:

Table 1. Comparison of D-S Theory and Probability Theory.

| D-S Theory | Probability Theory |
|---|---|
| m(U) does not have to equal 1 | $\sum_j P_j = 1$ |
| If $X \subseteq Y$, m(X) ≤ m(Y) is not necessary | If $X \subseteq Y$, P(X) ≤ P(Y) is necessary |
| there is no relation between m(X) and m(−X) | P(X) + P(¬X) = 1 |

Next is the concept of belief function. Belief Function (*Bel*) is the total trust of a set and all its subsets and is defined as following:

$$Bel(X) = \sum_{Y \subseteq X} m(Y) \tag{11}$$

Mass is trust about a set, and not any of its subsets. Mass is a more local trust, whereas the belief function applies to a set and any subset of that set, and *Bel* is a more global trust. The plausibility function is also a basic concept of D-S theory which denotes the confidence of not denying the estimated charging load result. In our approach, the term A denotes one prediction result of the EV charging load. It is the sum of the underlying probability assignments for each subset that intersects A. For instance, in our three loading predictions A, B, and C, all subsets intersecting the prediction A would be {A}, {A, B}, {A, C} and {A, B, C}, meaning that $Bel(X)$ would be the sum of the masses of the subsets {A}, {A, B}, {A, C} and {A, B, C}. $Bel(\overline{X})$ represents subsets excluding all subsets intersecting with A. In this study, $Bel(\overline{X})$ corresponds to {B}, {C}, and {B, C}. *Pls(X)*, read as plausibility of "X" is given by:

$$Pls(X) = 1 - Bel(\overline{X}) = 1 - \sum_{Y \subseteq \overline{X}} m(Y) \tag{12}$$

The overall procedure of the D-S theory task, which is employed to find the optimal charging loads forecasting, is demonstrated in Algorithm 2.

**Algorithm 2.** The Proposed D-S Theory Algorithm Pseudocode.

input: *Frames of Discernment U* = {A,B,C}
output: *Optimal Result*
1: *n = length (U)*
2:　function MassFunction(*U*)
3:　　$2^U$ = {*all subnets of U* }
4:　　$X = \forall 2^U$
5:　　if 　∑m(X) = 1 then m: $2^U \rightarrow$ [0, 1]
6:　　end if
7:　　return $m(2^U)$
8: end function
9:
10:　function SynthesisRule($X = \forall 2^U$, $m_1(X)$, $m_2(X)$)
11:　　*SynthesisMass* = 0
12:　　*K* = 0
13:　　$A = \forall 2^U$
14:　　$B = \forall 2^U$
15:　　if $A \cap B = \emptyset$ then
16:　　$K = K + m_1(A) * m_2(B)$
17:　　end if
18:　　if $A \cap B = X$ then
19:　　*SynthesisMass* = *SynthesisMass* + $m_1(A) * m_2(B)$
20:　　end if

```
21:    SynthesisMass = SynthesisMass/(1 − K)
22:    return SynthesisMass
23: end function
24:
25:    function ConfidenceInterval (X = ∀2^U)
26:    A = ∀2^U
27:    Bel(X) = 0
28:    Pls(X)  = 0
29:    if A ∈ X then
30:    Bel(X) = Bel(X) + m(A)
31:    else
32:    Pls(X) = Pls(X) + m(A)
33:    end if
34:    Pls(X) = 1 − Pls(X)
35:    CI(X) = [Bel(X), Pls(X)]
36:    return CI(X)
37: end function
```

## 3. System Implementation

In this section, the approach to implementing the actual model is discussed. First, we define the various inputs to the model and then the implementation of the DS-theory technique.

Figure 2 shows an overview of how the model is implemented. The description of the various input parameters is also presented in Figure 3. The various input parameters (1, 2, and 3) represent the following samples:

I. (1) represents the sample load data of the last moment and the parameters of that moment used to predict the current load.
II. (2) represents the sample load data of the last moment and the parameters of the current moment used to predict the current load.
III. (3) represents the sample load data of the last several moments and the corresponding parameters of the last several moments used to predict the current load.

These three samples serve as the input for the developed LSTM model. The output of the LSTM model now becomes the input of the data fusion model for the purpose of optimizing the initial predictions from the LSTM model. To implement the D-S theory technique, the predictions of the LSTM model with different input features based on different information of previous days is used. The results of these three predictions are V1, V2 and V3, respectively. D-S theory selects the appropriate course of action based on the output from the LSTM model to produce the most accurate forecast of the loads for the next 24-hour time horizon. To verify the credibility of predictions of V1, V2 and V3, the previous realistic charging loads are employed to show the accuracy of predictions. The predictions of V1, V2 and V3 in last 5 h, last 5 h to 10 h and last 10 h to 15 h are employed for this comparison purpose. The comparison results between the predictions and real (actual) loads are the events (E1, E2 and E3).

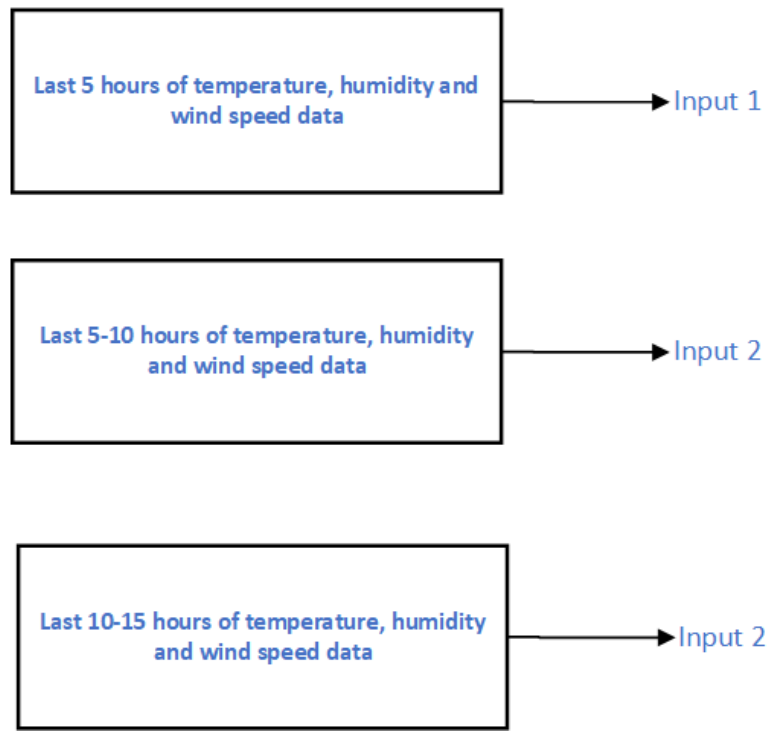

**Figure 3.** Description of various input parameters.

The mass (quality) function of the LSTM model predictions with different input features based on the definition of D-S theory are given by:

$$M(E_i V_j) = \frac{\sum_k^5 100 - \frac{\left(Ld_{E_i V_j} - Ld_{reality_k}\right)}{Ld_{reality_k}} \times 100}{5} \quad (13)$$

Based on the mass function of $E_i V_j$ and D-S theory, the decision matrix for combining the forecasting results of the events E1 and E2 results as shown in Tables 2 and 3.

**Table 2.** Decision Matrix of Combining Event 1 and Event 2 Based on D-S Theory.

|  | $E_1 V_1$ 30% | $E_1 V_2$ 26% | $E_1 V_3$ 44% |
|---|---|---|---|
| $E_2 V_1$ 31% | $V_1$ 9.3% | $V_1 V_2$ 8.06% | $V_1 V_3$ 13.64% |
| $E_2 V_2$ 34% | $V_1 V_2$ 10.2% | $V_2$ 8.84% | $V_2 V_3$ 14.95% |
| $E_2 V_3$ 35% | $V_1 V_3$ 10.5% | $V_2 V_3$ 9.1% | $V_3$ 15.4% |

Table 3. Decision Matrix of Combining Event 2 and Event 3 Based on D-S Theory.

|  | $V_1$ | $V_2$ | $V_3$ | $V_1V_2$ | $V_1V_3$ | $V_2V_3$ |
|---|---|---|---|---|---|---|
|  | 9.3% | 8.84% | 15.4% | 18.26% | 24.14% | 24.05% |
| $E_3V_1$ | $V_1$ | $V_1V_2$ | $V_1V_3$ | $V_1V_2$ | $V_1V_3$ | $V_1V_2V_3$ |
| **24%** | 2.23% | 2.12% | 3.70% | 4.38% | 5.79% | 5.77% |
| $E_3V_2$ | $V_1V_2$ | $V_2$ | $V_2V_3$ | $V_1V_2$ | $V_1V_2V_3$ | $V_2V_3$ |
| **41%** | 3.81% | 3.62% | 6.31% | 7.49% | 9.90% | 9.86% |
| $E_3V_3$ | $V_1V_3$ | $V_2V_3$ | $V_3$ | $V_1V_2V_3$ | $V_1V_3$ | $V_2V_3$ |
| **35%** | 3.26% | 3.09% | 5.39% | 6.39% | 8.45% | 8.42% |
| $V_1$ | $V_2$ | $V_3$ | $V_1V_2$ | $V_1V_3$ | $V_2V_3$ | $V_1V_2V_3$ |
| **2.23%** | 3.63% | 5.39% | 17.80% | 21.20% | 27.68% | 22.06% |

## 4. Simulation Results and Discussion

Our research proposes a novel method for forecasting EV charging load (energy demand). The forecast of charging load of EVs is very important as it allows EV charging stations to plan to meet energy requirements. First, a multi-input LSTM model was modeled to take three different forecast parameters for prediction of the charging load required. The input forecasting parameters consisted of data on weather conditions such as temperature, humidity and wind speed. These data were sourced from the UCI database [50]. The data was then put in a format that the algorithm could process to generate results. The values of the temperature, humidity, wind speed, and the power were normalized on a scale of 0–1 to make the data easier to handle, and the weight of the input parameters were set as the same. The model was then implemented using Python 3.8. The results from the initial prediction from the LSTM model for the various input parameters are shown in Figures 4–6. Descriptions of the various inputs utilized for the prediction are shown in Figure 3 and described in Section 3. Figures 4–6 show the forecasts with inputs (1), (2) and (3) respectively. To improve further upon this prediction, the data fusion model (DS-Theory) was used to enhance the predictions in Figures 4–6. A comparison between the forecast of the optimized model through data fusion and the various inputs of the LSTM model is presented in Figure 7. The performance of the created model was assessed using the widely used performance indicator, mean absolute error (MAE). The MAE was calculated as [51]:

$$MAE = \frac{1}{N}\sum|\hat{y}_i - y_i| \quad (14)$$

where $N$ is the sample number, $\hat{y}_i$ is the forecasted load demand and $y_i$ is the actual load demand.

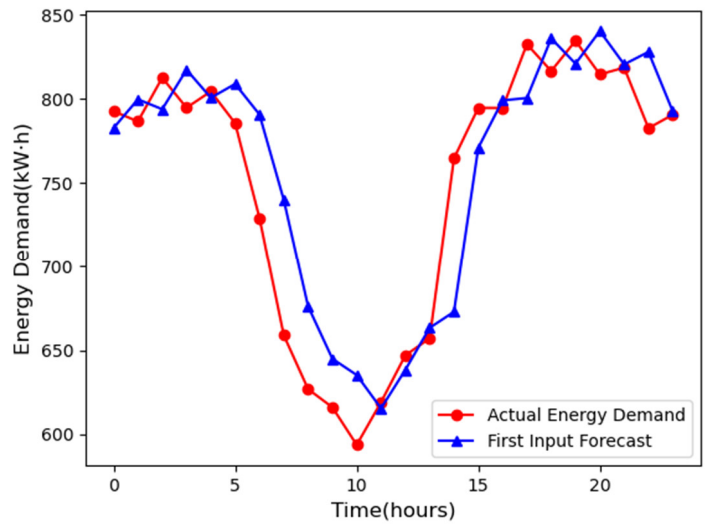

**Figure 4.** Forecast using only the first input parameter of the LSTM model.

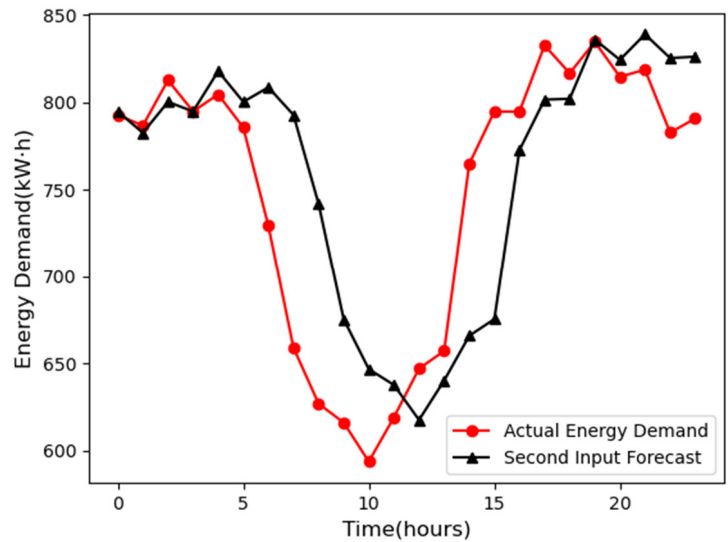

**Figure 5.** Forecast using only the second input parameter of the LSTM model.

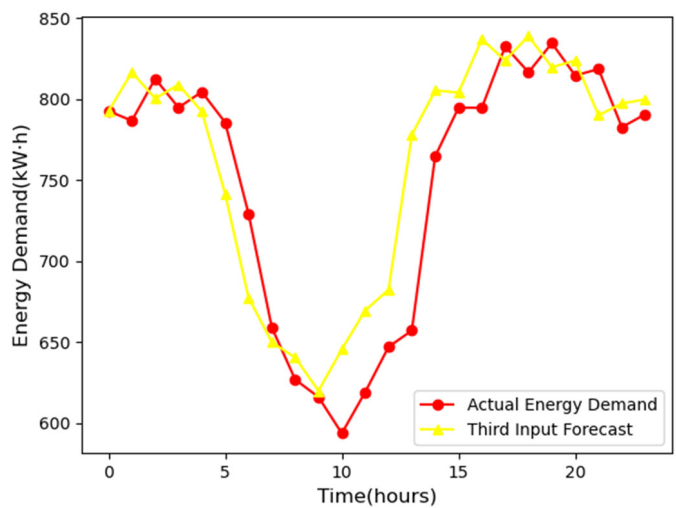

**Figure 6.** Forecast using only the third input parameter of the LSTM model.

The optimized prediction is shown in Figure 8. To compare the performance of the LSTM predictions (using the various input parameters) and the optimized prediction by the data fusion (DS theory) model, the relative mean squared error was found for all the prediction models. Figure 7 shows a graphical comparison between the various models and the actual energy demand of the EVs from the charging stations. Figure 9 shows a comparison of the prediction errors of the various prediction models (the LSTM and optimized data fusion). From Figure 9, the proposed model has a prediction error of 3.29%, which is better than other prediction models.

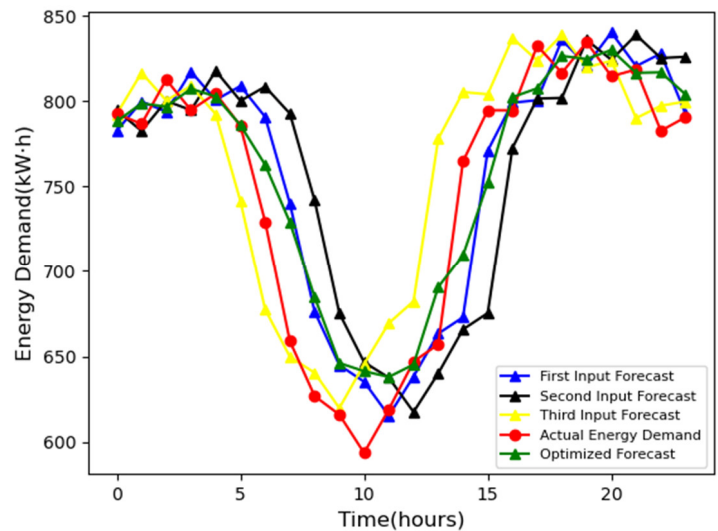

**Figure 7.** Comparison of LSTM results, optimized results and actual energy demand.

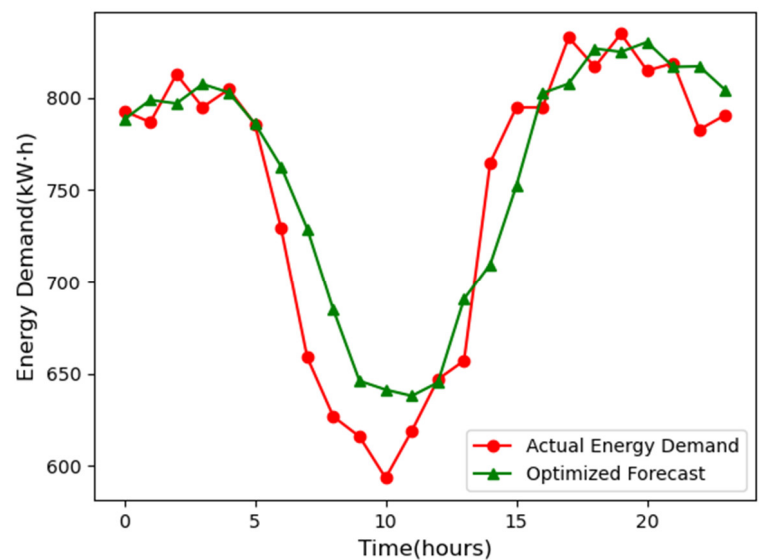

**Figure 8.** Results of the optimized forecast model.

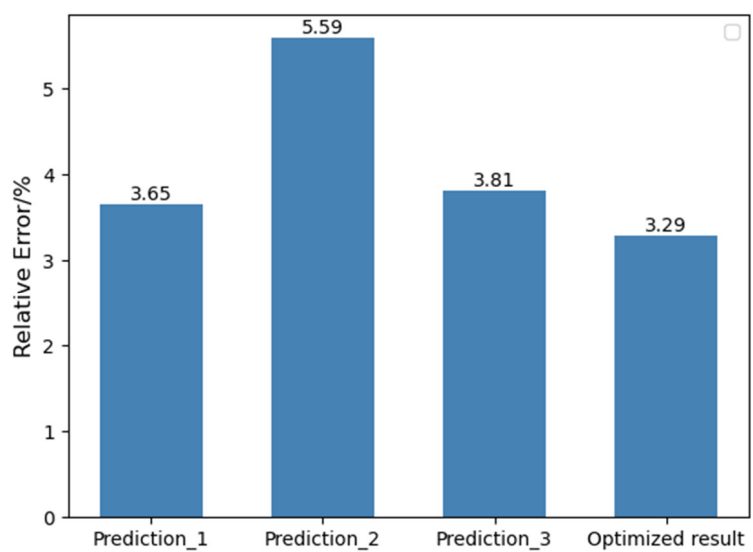

**Figure 9.** Error comparison between the optimized model and other prediction models.

## 5. Conclusions

To increase the precision of forecasts of the energy demand of electric vehicles using a deep learning model, we used a forecasting approach based on multi-feature data fusion. First, an initial prediction model based on LSTM was developed and implemented. Using data on weather conditions and charging load demands from UCI, the LSTM model was used to make three different initial predictions. The proposed DS-theory model was then used to optimize the initial predictions generated by our multi-input LSTM model. The proposed DS-theory model had a relative error of 3.29% which was an improvement on predictions generated by the LSTM model alone. This e proves the viability of the proposed model in improving EV load forecasting, which is important in planning for charging stations.

**Author Contributions:** Conceptualization, P.A., Z.Z. and A.S.A-S; methodology, Z.Z. and P.A.; software, Z.Z.; validation, Z.Z., P.A. and A.S.A.-S.; formal analysis, P.A., Z.Z. and A.S.A.-S.; investigation, Z.Z., P.A. and A.S.A.-S.; resources, A.S.A.-S.; data curation, Z.Z. and P.A.; writing—original draft preparation, P.A. and Z.Z.; writing—review and editing, P.A., Z.Z. and A.S.A-S; visualization, P.A., Z.Z. and A.S.A.-S.; supervision, A.S.A.-S. All authors have read and agreed to the published version of the manuscript.

**Funding:** This work is supported by ASPIRE-ViP project.

**Acknowledgments:** This is to acknowledge the value of NEP 3.0 in supporting the leadership of the third author.

**Conflicts of Interest:** The authors declare no conflict of interest.